\documentclass[preprint,aps]{revtex4}
\begin{document}
\title{Spatial entanglement of twin quantum images}
\author{Enrico Brambilla, Alessandra Gatti, Patrick Navez and 
Luigi A. Lugiato}
\affiliation{Istituto Nazionale per la  Fisica della Materia,
Dipartimento di Scienze CC FF MM, Universit\`a dell'Insubria,
via Valleggio 11, 22100 COMO, Italy}
\date{\today}
\begin{abstract}
We show that spatial entanglement of two twin images obtained by
parametric down-conversion is complete, i.e. concerns both
amplitude and phase. This is realised through a
homodyne detection of these images which allows for measurement of
the field quadrature components. EPR correlations are 
shown to exist between symmetrical pixels   
of the two images.
The best possible
correlation is obtained by adjusting the phase of the local
oscillator field (LO) in the area of maximal amplification.
The results for quadrature components hold unchanged even 
in absence of any input image i.e. for pure parametric 
fluorescence. In this case they are not related to 
intensity and phase fluctuations. 
\end{abstract}
\maketitle
\section{Introduction}

Optical systems which display quantum entanglement properties in the spatial
domain are of great interest for applications, 
since the amount of information that can be manipulated and processed
in parallel exploiting quantum correlation effects increases 
substancially with respect to the case of single mode beams.
Recently, there has been a rise of interest in the 
utilisation of entangled beam in optical imaging 
(quantum imaging) \cite{Jost,Images,Fabre,Images2}.

In this paper, we consider the field generated through the process of
frequency down-conversion
in a travelling wave optical parametric amplifier (OPA).
In \cite{Images,Images2} it was  demonstrated that such a system,
when coupled with an appropriate
classical imaging device, is able to generate
two symmetrical 
amplified copies of an injected input image
that are strongly correlated one to each other:
they indeed display synchronized local intensity fluctuations at the 
level of quantum noise, and for this reason they may be referred
to as twin images.

Here we present new results, that consolidate and complete the 
picture, showing that the two output images
are locally correlated, not only with respect to intensity fluctuations,
but also to ``phase'' fluctuations. 
To carry out a phase-sensitive measurements, 
we consider a homodyne detection scheme that allows us
to compare the fluctuations
of field quadrature components from two corresponding 
pixels of the two output images. 
We find in general that, for an arbitrary quadrature 
component characterized by the phase $\phi_L$ of 
the local oscillator, the difference between the 
fluctuations measured in two symmetrical pixels 
displays exactly the same spectrum as the sum of 
the fluctuations in the orthogonal quadrature 
component $\phi_L + \pi/2$.
The common value can be reduced 
well below the shot noise level over the whole image 
area, provided that the amplification is large enough 
and the phase $\phi_L$ is correctly adjusted.
The choice of the phase is crucial to obtain a large 
level of quantum correlations between symmetrical pixels 
for the quadrature component $\phi_L$ and 
anticorrelation for the quadrature component 
$\phi_L + \pi/2$. 
Thus, the twin images exhibit a complete 
Einstein-Podolsky-Rosen (EPR) \cite{epr35} 
entanglement with respect to continous 
variables \cite{epr92}. 

Since quantum correlations are shown to exist for any couple of  
symmetrical pixels over the whole area of the output images,   
we speak of spatial quantum entanglement.
The system exhibits a {\em spatial} realization of
the EPR paradox for 
two orthogonal quadrature components
of the output field similar to that shown in \cite{epr} 
for the case of the parametric oscillator below 
threshold. In comparison with the analysis of 
\cite{epr}, we consider here also the case in which 
a coherent image is injected into the system. 
Furthermore, the consideration of an OPA, instead 
of an optical parametric oscillator with 
spherical mirrors, allows for obtaining here 
completely analytical results.

In presence of an input image, the mean output field is 
different from zero, and therefore amplitude 
(i.e. intensity) and phase fluctuations correspond 
to special cases of quadrature fluctuations.  
Therefore the previous analysis allows for concluding 
in a rather straightforward way that in symmetrical 
pixels not only quantum intensity fluctuations are 
strongly correlated, but phase fluctuations are 
anticorrelated in the same amount.

The paper is divided as follows. After a presentation
of the optical image amplification scheme in section II,
in the third section we study the fluctuation spectrum
of the quadrature components
measured in homodyne detection.
The fourth section is devoted to the discussion of 
amplitude and phase fluctuations. 
The final section
includes conclusions and perspectives.

\section{Optical image amplification scheme}

The experimental procedure to
generate a pair of quantum entangled
images through the process of
parametric down-conversion close to the degenerate frequency
has been discussed in previous papers \cite{Images,Images2} and 
can be summarized as follows. The $\chi^{(2)}$ crystal is enclosed
between two lenses $L$ and $L'$, as shown in Fig.I. 
We take the $z$ axis as the main light propagation 
direction and indicate with $\vec{x}=(x,y)$
the point coordinates in a generic transverse plane.
Not shown in the figure is the coherent pump field that activates
the process of down-conversion and which we take as an ideal
classical monochromatic plane wave
of frequency $\omega_p$, propagating inside the crystal along the $z$
direction.\\
The crystal slab of width $l_c$, ideally infinite in the transverse
directions, is cut for type I quasi-collinear 
phase-matching at the degenerate
frequency $\omega_p/2$. Under these assumptions,
each elementary down-conversion
process corresponds to the splitting of a pump photon of frequency
$\omega_p$ into a pair of photons of frequencies $\omega_p/2+\Omega$
and $\omega_p/2-\Omega$ (with $\Omega\ll\omega_p$),
propagating with the same polarization and with opposite transverse
wavevectors $\vec{q}$ and $-\vec{q}$.

We designate by $a_1(\vec{x},t)$, $a_2(\vec{x},t)$, $a_3(\vec{x},t)$, $a_4(\vec{x},t)$,  
the slowly varying envelope operators of the down-converted field 
(with respect to the carrier frequency $\omega_p/2$) in the input plane
$P_1$, the entrance plane $P_2$ of the $\chi^{(2)}$ crystal slab,
its exit plane $P_3$, and the output plane $P_4$, respectively 
(Fig. 1).
We denote by $a_i(\vec{x},\Omega)$, $a_i(\vec{q},\Omega)~~(i=1,..4)$
their Fourier transforms in time and in space-time respectivelly.
The purpose of the two lenses is to map the Fourier plane $(q_x,q_y)$
into the physical plane $(x,y)$.
In this manner, if an optical image is injected at the
degenerate frequency $\omega_p/2$ in the object plane $P_1$,
the system amplifies portions of this image rather
than a band of its $q$-vectors.
Indeed, the input-output transformation which describes propagation
inside the crystal in the linear regime, assuming that 
pump depletion and losses are negligible, can be written
as \cite{kolobov99}:
\begin{eqnarray}
\label{inputoutput1}
a_3(\vec{q},\Omega )
=u(\vec{q},\Omega )a_2(\vec{q},\Omega )+
 v(\vec{q},\Omega ) a_2^{\dag }(-\vec{q},-\Omega ) \: ,
\label{eq5}
\end{eqnarray}
The presence of the lenses converts it into a
relation between the real-space field operators
in the plane $P_1$ and $P_4$:
\begin{eqnarray}
\label{inputoutput2}
a_4(\vec{x},\Omega )
=\overline{u}(\vec{x},\Omega) a_1(-\vec{x},\Omega )+
\overline{v}(\vec{x},\Omega) a_1^{\dag}(\vec{x},-\Omega ) \: ,
\label{eq6}
\end{eqnarray}
where
\begin{eqnarray}
\label{uvbarrati}
\overline{u}(\vec{x},\Omega)
=-u\left(\frac{2\pi \vec{x}}{\lambda f},\Omega\right), \;
\overline{v}(\vec{x},\Omega)
=v\left(\frac{2\pi \vec{x} }{\lambda f},\Omega\right) \;,
\label{eq7}
\end{eqnarray}
$f$ is the focal length of the two lenses and $\lambda$ is the wavelength
of the down-converted field.
The explicit expressions of the gain coefficients $u(\vec{q},\Omega)$
and $v(\vec{q},\Omega)$ can be found in \cite{kolobov89}.
Here we just notice that they depend on the linear gain
parameter $\sigma$ and on the dispersion properties of the crystal; they  
are functions of the modulus of $\vec{q}$ and $\Omega$ 
and, for $\Omega=0$,
display a broad maximum, 
corresponding to that transverse wavenumber which is phase-matched
at the degeneracy frequency, for which (within the paraxial approximation)
\begin{eqnarray}\label{q2}
q^2 = k_s^2-(k_p/2)^2\approx k_s \Delta_0\;,
\end{eqnarray}
where $k_s$, $k_p$ are the wave numbers
of the signal and the pump field at the carrier frequencies 
$\omega_p/2$ and $\omega_p$, respectivelly, 
and
$\Delta_0=2k_s-k_p$ is the collinear phase mismatch parameter 
which is assumed non negative.
The width of the {\em plateau} around the value (\ref{q2}) 
is on the order of $q_0=\sqrt{k_s/l_c}$,
the variation scale of $|u|$ and $|v|$ in the spatial frequency domain.

We underline that all the results that
follow do not depend on the particular form of the gain
functions, but rely on the fact that they satisfy the following
unitarity conditions
\begin{eqnarray}
\label{unitarity}
|u(\vec{q},\Omega)|^2-|v(\vec{q},\Omega)|^2&=&1, \;\\
u(\vec{q},\Omega)v(-\vec{q},-\Omega)&=&
u(-\vec{q},-\Omega)v(\vec{q},\Omega),\;\nonumber
\end{eqnarray}
which guarantee that the free field commutation rules 
are preserved:
\begin{eqnarray}
\label{commut}
\left[a_i(\vec{q},\Omega),a_i^{\dag}(\vec{q}',\Omega')\right]
&=&\delta(\vec{q}-\vec{q}')\delta(\Omega-\Omega'),\;\\
\left[a_i(\vec{q},t),a_i(\vec{q}',t')\right]&=&0.\;\nonumber
~~~~~~~~(i=1,2,3,4)
\end{eqnarray}
On the other hand, with respect to other systems which 
exhibit input/output relations of the same form 
(e.g. optical parametric oscillators, see e.g. \cite{kolobov95}), 
the large spatial bandwidth $q_0$ of the amplifier 
makes this travelling-wave scheme a good candidate for high resolution
image amplification. 

For the scheme of Fig.1, the region in the transverse plane
which can be efficiently amplified without distortion
has a linear size on the order of
\begin{equation}
x_0=\frac{\lambda f}{2\pi}q_0\;,
\end{equation}
which represents the width of the {\em plateau} of the real-space gain functions
(\ref{uvbarrati}). Such a region
has either the shape of a disc of area $\sim S_0=x_0^2$ 
centered at the origin, or a
ring of width $\sim x_0$, depending on the possibility to have
collinear ($\Delta_0=0$) or non-collinear ($\Delta_0>0$)
phase-matching at $\Omega=0$, respectivelly. 
We assume that the input image
is a coherent stationary field of frequency $\omega_p/2$ 
confined in this region of plane $P_1$ (see Fig.1) so that
\begin{equation}\label{a1}
\langle  a_1(\vec{x},\Omega) \rangle
=\sqrt{2 \pi} \delta (\Omega) \alpha_{in} (\vec{x}) \; .
\end{equation}
As explained in details in \cite{Images2,kolobov99}, whenever the input image 
is symmetric with respect to the system axis, the device
works as a phase-sensitive amplifier (see in this 
connection also \cite{Levenson}). 
In this case, the phase of the input image must be selected
in order to optimize the gain.
Assuming the input image is duplicated before amplification
by means of a classical imaging device which allows to obtain
a symmetrical field  ditribution 
(i.e. $\alpha_{in}(-\vec{x})=\alpha_{in}(\vec{x})$),
the system is able to generate in the output plane
two amplified copies 
that are far better correlated in space-time than the originals,
meaning by this that they display perfectly 
(in the ideal case) synchronized local
intensity fluctuations. It was also demonstrated \cite{Images2} 
that in the limit of
high gain, the signal-to-noise ratio as measured from a small portion of the
input image before duplication is preserved in the corresponding
portions of the two output images:
{\em noiseless} amplification is therefore achieved
for both output channels taken separately 
(see \cite{Kumar} for an experimental observation 
of noiseless amplification of images).

In \cite{Images} an alternative way to generate a pair
of quantum correlated images (also called {\em twin images}) was considered; it
consists in injecting a single input image
asymetrically, for example by confining it
to the upper half of plane $P_1$ as shown in Fig.1.
This configuration does not
require a duplication system and
presents the further advantage that the gain does not depend on the phase
of the input field because the systems works 
as a phase-insensitive amplifier.
However, the fidelity with which information is transferred
is worse than in the phase-sensitive case, since the signal-to-noise ratio
is deteriorated at least by a factor two in the amplification
process (a feature common to all phase-insensitive optical amplifiers
\cite{caves}).

Most of the results presented in this paper do not depend on the particular
injection scheme, so no assumption are made on the input intensity
distribution $|\alpha_{in}(\vec{x})|^2$.
Imperfect detection can be modelled in the usual way,
by coupling the output field operator $a_4(\vec{x},t)$
with an independent operator field
$a_N(\vec{x},t)$ which acts on the vacuum state.
The contribution $a_N$ 
describes the noise added by losses 
in the detection process; thus the effective output field 
measured by a detector of quantum efficiency $\eta\leq 1$ is
\begin{equation}\label{aD}
a_D(\vec{x},t)=
\sqrt{\eta}a_4(\vec{x},t)+ \sqrt{1-\eta}a_N(\vec{x},t)
\end{equation}
and the corresponding photon flux density is
\begin{equation}\label{i}
i(\vec{x},t) =
a^\dagger_D(\vec{x},t)a_D(\vec{x},t) \; .
\end{equation}
As shown in Fig.1,
at the exit face of the crystal we insert a pupil
of area $S_p$, an element 
that allows for eliminating divergencies which arise in the
calculation of the field mean intensity and correlation functions
when dealing with a system of infinite transverse dimensions 
\cite{kolobov99,kolobov95}.
It also determines
the characteristic resolution area of the device
in the detection plane, which is $S_R=(\lambda f)^2/S_p$.
This finite size optical element introduces a convolution integral with
the pupil response function in the r.h.s. of Eq.~(\ref{inputoutput2})
and, as a consequence, the points of the input image are
spread into diffraction spots of area $S_R$ in the output image.
However, analytical calculations are performed in the
{\em low diffraction limit},
assuming that the diffraction spot size $\sqrt{S_R}$
is much smaller than both $x_0$ and the variation scale of the input 
image intensity.  
Considering a single pixel detector (labelled by index $j$)
that intercepts the photons arriving on an area $R_j$
which is large in comparison with 
$S_R$, the mean value of the measured
photocurrent is then \cite{Images2}:
\begin{eqnarray}
\label{intensity}
\langle i_j (t) \rangle &=& \int_{R_j} \!\!\!d\vec{x}\,
\langle i(\vec{x},t)\rangle \nonumber\\
&=&
\eta\int_{R_j}\!\!\!d\vec{x}\,
|\overline{u}(\vec{x},0)\alpha_{in}(-\vec{x})+\overline{v}(\vec{x},0)\alpha_{in}^*(\vec{x})|^2
+\frac{\eta}{S_R}
\int_{R_j} \!\!\!d\vec{x}\,
\int_{-\infty}^\infty \frac{d\Omega'}{2\pi}
|\overline{v}(\vec{x},\Omega')|^2 \; .
\end{eqnarray}
The first integral represents the amplified coherent input field
while the second integral is the contribution
coming from spontaneous parametric down-conversion.
The ratio $S_0/S_R$ gives an evaluation
of the number of details of the input image which
can be resolved in the detection plane (e.g. with the pixel
array of a CCD camera). Moreover, quantum
correlation effects tends to disappear when $S_R\rightarrow S_0$,
since in this limit the signal and idler photons of each
down-converted pair can no more be resolved separately,
because of the large diffraction spread in $q$-space.
Making $S_R$ as small as possible with respect to $S_0$
is therefore a necessary requirement that must be taken into
account in experiments.
However, this leads to an increase of the
spontaneous emission contribution which goes at the expense
of the visibility of the amplified input image.
This last circumstance imposes a lower limit on the intensity of the
input image (see \cite{Images,kolobov99,kolobov95} for more details).

\section{Correlations measurement in a homodyne detection scheme}
A homodyne detection scheme allows for the measurement of
a particular quadrature component of the field.
It consists in a beam splitter that combines the output field
with a coherent field of much higher intensity, $\alpha_L(\vec{x})$,
which can be treated as a classical quantity and is usually referred to as
the local oscillator field (LO). 
In the balanced version a 50/50 beam splitter is used,
so that the operators associated to the fields
coming from the two output ports of the beam-splitter , labelled by $b$ and $c$, are
\begin{eqnarray}
a^{b,c}(\vec{x},t)=\left[a_4(\vec{x},t) \pm \alpha_L(\vec{x})\right]/\sqrt{2}
\end{eqnarray}
and the effective fields seen by two
identical detectors of quantum efficiency $\eta$ 
in the two ports $b$ and $c$ are
\begin{eqnarray}
a_{D}^{b,c}(\vec{x},t)=
\sqrt{\eta} ~a^{b,c}(\vec{x},t) +
\sqrt{1-\eta} ~a^{b,c}_N(\vec{x},t)\;,
\end{eqnarray}
where $a_{N}^{b,c}(\vec{x},t)$ describe the noise added
in the detection process. When the corresponding intensities
are electronically substracted, one obtains a direct measure
of the quadrature component of the output field $a_4$ selected by the phase
of the LO, more precisely
\begin{eqnarray}
Z_{\phi_L} (\vec{x},t)&=&
a^{b\dagger}_{D}(\vec{x},t) a^b_{D}(\vec{x},t) -
a^{c\dagger}_{D}(\vec{x},t) a^c_{D}(\vec{x},t)\\
& &\stackrel{\eta \to 1} {\longrightarrow}     
\rho_L(\vec{x}) \left[ a^\dagger_4(\vec{x},t) e^{i\phi_L(\vec{x})}
                     + a_4(\vec{x},t) e^{-i\phi_L(\vec{x})}\right]\;,
\end{eqnarray}
where $\rho_L(\vec{x})=|\alpha_L(\vec{x})|$ and
$\phi_L(\vec{x})=\arg\alpha_L(\vec{x})$.
Taking into account the finite size of the pixel detection area $R_j$,
the measured quantity is
\begin{equation}\label{Zx}
Z^{(j)}_{\phi_L}(t)=\int_{R_j}~~d \vec{x} Z_{\phi_L}(\vec{x},t)\; .
\end{equation}
We now want to compare the fluctuations of the field quadrature
measured in two symmetrical pixels $j=1$ and $j=2$ 
of the signal and idler image. To this aim, we 
consider the sum and the difference of the quadrature obtained
from two symmetrical detection regions $R_1$ and $R_2$:
\begin{equation}\label{Zxpm}
Z_{\phi_L}^{(\pm)}(t)=Z^{(1)}_{\phi_L}(t)\pm Z^{(2)}_{\phi_L}(t)\;.
\end{equation}
The corresponding fluctuation spectra, defined as
\begin{eqnarray} \label{defV}
& &V^{(\pm)}_{\phi_L}(\Omega)=\int_{-\infty}^{\infty}dt \,
e^{i\Omega t}\langle \delta Z_{\phi_L}^{(\pm)}(t)
                     \delta Z_{\phi_L}^{(\pm)}(0)\rangle\;,\\
& &\delta Z_{\phi_L}^{(\pm)}(t)
= Z_{\phi_L}^{(\pm)}(t)-\langle Z_{\phi_L}^{(\pm)}(t)\rangle\;,\nonumber
\end{eqnarray}
describe the degree of correlation between the observables
$Z^{(1)}_{\phi_L}$ and $Z^{(2)}_{\phi_L}$.
Using the input-output transformation (\ref{inputoutput2}),
the commutation rules (\ref{commut}) and the fact that the 
input image is coherent we obtain the following
relations:
\begin{mathletters}
\begin{eqnarray}
V^{(-)}_{\phi_L}(\Omega)&=&V^{(+)}_{\phi_L+\pi/2}(\Omega)
\nonumber \\
&=&
\label{vpm}
(1-\eta)(SN)_{LO}
%+\eta\langle i_+(t)\rangle
+ \eta^2 \int_{R_1+R_2}\!\!\!\! d\vec{x}
|\overline{u}(\vec{x},\Omega)\alpha_L^*(\vec{x})
-\overline{v}^*(\vec{x},\Omega)\alpha_L(-\vec{x})|^2 \; ,
\end{eqnarray}
\end{mathletters}
where
\begin{equation}\label{SNLO}
(SN)_{LO}=\eta\int_{R_1+R_2}\!\!\!\!d\vec{x}~\rho_L^2(\vec{x})
\end{equation}
is the shot noise level determined by the LO on the two detectors
(we assumed $|\alpha_L(\vec{x})|^2 \gg \langle j(\vec{x},t)\rangle$).
Next, we assume that the LO is symmetric with respect to the 
system axis, i.e. $\alpha_L(\vec{x})=\alpha_L(-\vec{x})$. 
Because $\overline{v}(\vec{x},\Omega)= \overline{v}(\vec{x},-\Omega)$
we can write:
\begin{eqnarray}
\label{vpm2}
V^{(-)}_{\phi_L}(\Omega)&=&V^{(+)}_{\phi_L+\pi/2}(\Omega)
=
(1-\eta)(SN)_{LO}
+ \eta^2 \int_{R_1+R_2}\!\!\!\! d\vec{x}
F(\vec{x},\Omega)\rho_L^2(\vec{x})
\end{eqnarray}
where
\begin{equation}
\label{g}
F(\vec{x},\Omega)=|\overline{u}(\vec{x},\Omega)e^{-i\phi_L(\vec{x})}
           -\overline{v}^*(\vec{x},\Omega)e^{i\phi_L(\vec{x})}|^2 \;.
\end{equation}
Note that the above expression 
corresponds to the fluctuation spectrum 
normalized to shot noise when $\eta=1$ and the pixel  
area is small with respect to $x_0^2$ and to the square 
of the scale of variation of $\alpha(\vec{x})$. 
In this case $\vec{x}$ in Eq.(\ref{vpm2}) must be taken 
as the central point of pixel 1 or pixel 2; the result 
is the same for both pixels because 
$\overline{u}(\vec{x},\Omega)=
\overline{u}(-\vec{x},\Omega)$,$ 
\overline{v}(\vec{x},\Omega)=\overline{v}(\vec{x},\Omega)$
and $\phi_L(\vec{x})=\phi_L(-\vec{x})$. 

A first important result follows from the first
equality (\ref{vpm})), according to which $Z^{(1)}_{\phi_L}$
and $Z^{(2)}_{\phi_L}$ are correlated one to each other
exactly to the same extent as the corresponding
orthogonal quadrature components $Z^{(1)}_{\phi_L+\pi/2}$
and $Z^{(2)}_{\phi_L+\pi/2}$ are anti-correlated.
Second, the common fluctuation spectrum of the two
observables $Z^{(1)}_{\phi_L}-Z^{(2)}_{\phi_L}$ and
$Z^{(1)}_{\phi_L+\pi/2}+Z^{(2)}_{\phi_L+\pi/2}$
as given by expression (\ref{vpm}) does not
depend on the intensity and phase of the input image.
Hence the result is the same in the 
phase-insensitive and in the phase-sensitive scheme, and 
remains the same even in absence of an input image 
at all, i.e. in the case of pure parametric 
fluorescence.
Third, this spectrum can be reduced well below
the shot noise level, provided the gain is large enough
and the phase of the LO is correctly adjusted. Indeed, assuming that
\begin{equation}
\label{maxcorr}
\phi_L(\vec{x})=\frac{1}{2}\left(\arg\overline{u}(\vec{x},0)+
                    \arg\overline{v}(\vec{x},0)\right)=
\phi_{opt}(\vec{x})\;
\end{equation}
over the two detection areas, using the symmetry property
of the LO $\alpha_L(-\vec{x})=\alpha_L(\vec{x})$
and unitarity relations (\ref{unitarity}),
one obtains from Eq.~(\ref{vpm}) for $\Omega=0$:
\begin{eqnarray}
F(\vec{x},\Omega=0)=
 \frac{1}
  {\left[ |\overline{u}(\vec{x},0)|+|\overline{v}(\vec{x},0)|\right]^2}   \; ,
\end{eqnarray}
which goes to zero when
$|\overline{u}(\vec{x},0)|\sim|\overline{v}(\vec{x},0)|\gg 1$.
Under conditions of large gain and reasonably large
quantum efficiency, almost perfect correlation
between the selected quadratures can therefore be
obtained. 

It is interesting to relate the phase of optimum 
squeezing (in $Z^{(1)}_{\phi_L}-Z^{(2)}_{\phi_L}$) 
$\phi_{opt}$ for the LO with the phase of maximum 
amplification in the phase sensitive configuration.
The mean output field is in general
\begin{eqnarray}\label{aout}
\alpha_{out}(\vec{x})=
\langle a_4(\vec{x},t)\rangle=
\overline{u}(\vec{x},0)\alpha_{in}(-\vec{x})
+\overline{v}(\vec{x},0)\alpha^*_{in}(\vec{x}),
\end{eqnarray}
where we used Eq.(\ref{inputoutput2}) and Eq.(\ref{a1}). 
In the phase sensitive case $\alpha_{in}(-\vec{x})=
\alpha_{in}(\vec{x})$ we can write
\begin{eqnarray}
|\alpha_{out}(\vec{x})|^2=
G(\vec{x})|\alpha_{in}(\vec{x})|^2
\end{eqnarray}
with the phase-sensitive gain given by:
\begin{eqnarray}\label{Gain}
G(\vec{x})=|\overline{u}(\vec{x},0) e^{\phi_{in}(\vec{x})}
                     +\overline{v}(\vec{x},0) e^{-\phi_{in}(\vec{x})}|^2\;,
\end{eqnarray}
where $\phi_{in}(\vec{x})$ is the phase of 
$\alpha_{in}(\vec{x})$. We easily obtain that 
the maximum gain 
\begin{eqnarray}
G_{max}(\vec{x})=
||\overline{u}(\vec{x},0)|
+|\overline{v}(\vec{x},0)||^2\;
\end{eqnarray}
is obtained for:
\begin{equation}
\label{phiin}
\phi_{in}(\vec{x})=
\phi_{in}^{max}(\vec{x})=
\frac{1}{2}\left(\arg\overline{u}(\vec{x},0)-
                    \arg\overline{v}(\vec{x},0)\right)\;.
\end{equation}
On the other hand, from Eq.(\ref{aout}) with 
$\alpha_{in}(-\vec{x})=\alpha_{in}(\vec{x})$ one 
obtains that when $\phi_{in}=\phi_{in}^{max}$ the phase 
$\phi_{out}$ of the output field $\alpha_{out}$ 
is given by
\begin{equation}
\label{phiout}
\phi_{out}(\vec{x})=
\phi_{out}^{max}(\vec{x})=
\frac{1}{2}\left(\arg\overline{u}(\vec{x},0)+
                    \arg\overline{v}(\vec{x},0)\right)\;
\end{equation}
and therefore coincides with $\phi_{opt}(\vec{x})$ given 
by Eq.(\ref{maxcorr}). This leads to the following 
interpretation for $\phi_{opt}$: the phase $\phi_L$ of 
optimum squeezing in $Z_{\phi_L}^{(-)}$ coincides with the phase 
of the output field in the phase-sensitive configuration, 
provided the phase of the input field is selected to have 
maximal amplification. 
Note that in the special case of perfect phase matching one has 
$\arg\overline{u}(\vec{x},0)=0$ so that $\phi_{in}^{max}(\vec{x})$ 
given by Eq.(\ref{phiin}) coincides with the corresponding 
$\phi_{out}(\vec{x})$.

The results obtained for the observables $Z_{\phi_L}^{(\pm)}$
closely ressemble the situation of the
EPR paradox for continous variables demonstrated in \cite{epr92}, 
but generalised to many pixels (see also \cite{epr}) 
and to the presence of
input images. 
We notice indeed that the  conjugated observables
$X_j=\int_{-T_D/2}^{T_D/2}dt
Z_{\phi_L}^{(j)}(t)$ and
$P_j=\int_{-T_D/2}^{T_D/2}dt
Z_{\phi_L+\pi/2}^{(j)}(t)$ ($j=1,2$) obey the
uncertainty rule:
\begin{eqnarray}\label{uncertainty}
\langle \delta^2 X_j \rangle 
\langle \delta^2 P_j \rangle \geq \frac{1}{4}
\left[T_D\int_{R_1 +R_2}d\vec{x} \rho_L^2(\vec{x})\right]^2\; .
\end{eqnarray}
On the other hand, 
the following combination over
the two pixels: $X_-=X_1-X_2$ and $P_+=P_1+P_2$ are commuting
observables that can be simultanously determined.
When the time of measurement $T_D$ is much larger
than the inverse of the temporal bandwidth of the OPA, 
using Eq.(\ref{defV}), 
the uncertainty
of these observables can be directly related to the fluctuations
spectrum $V^{(-)}_{\phi_L}$ 
\begin{eqnarray}
\langle \delta^2 X_- \rangle=
\langle \delta^2 P_+ \rangle=
T_D V^{(-)}_{\phi_L}(\Omega=0)\;.
\end{eqnarray}

For $\eta=1$, an optimal adjustement of the LO phase allows these
uncertainties to reach almost a zero value for large
amplification and thus to display an apparent violation of the
Heisenberg rule:
\begin{eqnarray}\label{violation}
\langle \delta^2 X_- \rangle 
\langle \delta^2 P_+ \rangle < \frac{1}{4}
\left[T_D\int_{R_1+R_2}d\vec{x} \rho_L^2(\vec{x})\right]^2 \;.
\end{eqnarray}

However, it is impractical to synthetyze a LO with the phase 
variation prescribed by Eq.(\ref{maxcorr}). On the other hand, for a LO 
with constant phase, the condition (\ref{maxcorr})
concerning the phase of the LO can be exactly satisfied
only for a single couple of pixels of area small compared
to $S_0$, so that the gain functions $|\overline{u}(\vec{x},0)|$
and $|\overline{v}(\vec{x},0)|$ are nearly uniform over the
detection areas. We can however show that ,
by introducing an appropriate curvature the the wavefront of the LO
field, EPR-like correlations are present for each couple 
of symmetric pixels in the output over the whole gain 
region $S_0$. To this end, we allow the LO phase distribution 
to have a quadratic dependence on the spatial coordinate
(which corresponds to a spherical wavefront as one has e.g. in gaussian 
beams). The
wavefront curvature is selected in order to have the best
fit of the spatial dependance of $\phi_{opt}(\vec{x})$ in 
Eq.(\ref{maxcorr}).

Figure 2 plots the function $F(\vec{x},\Omega=0)$
in the limit where
$R_1$ and $R_2$ are small compared to $S_0$ and symmetric.
The collinear phase-mismatch at degeneracy is
$\Delta_0 l_c=0.5$ and the linear gain parameter is
$|\sigma|l_c=1.5$. 
%(See Ref.\cite{kolobov89} for
%the explicit expressions) 

Curve (a) corresponds to the ideal case, with $\phi_L$
satisfying condition (\ref{maxcorr}) everywhere in the
transverse plane,
and leads to a maximal amount of noise reduction
in the whole amplification region.
In curve (b), the phase of the LO is constant
and satisfies condition (\ref{maxcorr}) only in the
point of maximum gain $x_G$, where perfect phase-matching is
achieved, $|\overline{u}|^2\simeq 5.5$
and $F=(|\overline{u}|-|\overline{v}|)^2=\exp(-2|\sigma|l_c)
\simeq 0.1$.
Curve (c), obtained by optimizing the
phase with a quadratic term (i.e. we take the form
$\phi_L(\vec{x})= \Phi_0 + \Phi_2(|\vec{x}|-x_G)^2/x_0^2$ 
), is the best
that can be done with a gaussian LO and is close
to the ideal case.

\section{
Phase-intensity entanglement of the twin images}

Although the phase-sensitive measurement scheme considered in the last
section offers a picture of the spatial correlations
that can be observed in the output field, intensity
correlation measurements are more straightforward to
perform experimentally and lead also to interesting effects of
quantum noise reduction \cite{Images,Images2,fluor}.
The observable that displays reduced fluctuations with
respect to the coherent state level is the difference between
the direct photocurrents measured from two symmetrical
detection region $i_-=i_1-i_2$.
The corresponding fluctuation spectrum is 
\begin{eqnarray}
V_{i_-}(\Omega)
&=&
\int_{-\infty}^{\infty}
     dt~e^{i\Omega t}
       \langle \delta i_-(t)~\delta i_-(0) \rangle
\end{eqnarray}
By using (\ref{inputoutput2}), (\ref{commut}), (\ref{aD}), (\ref{i}), 
(\ref{intensity}) and the fact that input image is in a coherent 
state one obtains after lengthy but elementary calculations 
\begin{eqnarray}
V_{i_-}(\Omega)
&=&
(1-\eta)\langle i_+ \rangle\nonumber\\
& &+
\eta^2 \int_{R_1+R_2}\!\!\!\! d\vec{x}~
   |\overline{u}(\vec{x},\Omega)\alpha_{out}^*(\vec{x})
    -\overline{v}^*(\vec{x},-\Omega)\alpha_{out}(-\vec{x})|^2\nonumber\\
& &+\eta^2\frac{1}{S_R}\int_{R_1+R_2}\!\!\!\!d\vec{x}
     \int \frac{d\Omega'}{2\pi}\left[
     |\overline{u}(\vec{x},\Omega')|^2
     |\overline{v}(\vec{x},\Omega+\Omega')|^2\right.\nonumber\\
& &\hspace{3cm}\left.
   -\overline{u}(\vec{x},\Omega+\Omega')\overline{v}(\vec{x},-\Omega-\Omega')
    \overline{u}^*(\vec{x},\Omega')\overline{v}^*(\vec{x},-\Omega')
    \right]
						\label{ispectrum}
\end{eqnarray}
where $\alpha_{out}(\vec{x})$ is given by Eq.(\ref{aout}). 
The shot noise level corresponds to 
the photocurrent sum $\langle i_+ \rangle =
\langle i_1+i_2 \rangle$.
The second term on the l.h.s. of Eq.~(\ref{ispectrum})
arises from the interference 
between the amplified input field and the fluorescence field.
The last term, which does not depend on the presence of an input field,
is a pure noise contribution due
to the self-interference of the fluorescence field
and reduces to zero for $\Omega=0$ 
because $\overline{v}(\vec{x},-\Omega)=\overline{v}(\vec{x},\Omega)$ 
\cite{Images}.

Using the explicit expression of the
amplified input field (\ref{aout}) and the fact that 
$\overline{u}(\vec{x},\Omega)=\overline{u}(-\vec{x},\Omega)$, 
$\overline{v}(\vec{x},\Omega)=\overline{v}(-\vec{x},\Omega)$ 
and Eq.(\ref{unitarity}),
we find for
the zero frequency value of the spectrum
\begin{equation}
V_{i_-}(0)=(1-\eta)\langle i_+ \rangle
     +\eta^2 \int_{R_1+R_2}\!\!\!\! d\vec{x}~|\alpha_{in}(\vec{x})|^2
\end{equation}
As shown in \cite{Images} in the case of ideal detection ($\eta=1$),
the noise level of $i_-$ reduces therefore
to the noise of the input image over $R_1+R_2$. As a consequence,
under conditions of large gain, fluctuations are well below
the shot noise level.

It is important now to connect with the result for 
quadrature components obtained in the previous 
section. To this aim, let us first assume that the 
input field is strictly different from zero at least 
in some region of the transverse plane. Second, let us 
assume that the parametric values are such that the 
pure noise contribution in $V_{i_-}(\Omega)$ 
(i.e. the last term in Eq.(\ref{ispectrum})) is 
negligible and, similarly, that the second term 
in Eq.(\ref{intensity}) can be dropped. Thus 
expression (\ref{ispectrum})) reduces to
\begin{eqnarray}
V_{i_-}(\Omega)
&=&
(1-\eta)SN_- \nonumber\\
& &+
\eta^2 \int_{R_1+R_2}\!\!\!\! d\vec{x}~
   |\overline{u}(\vec{x},\Omega)\alpha_{out}^*(\vec{x})
    -\overline{v}^*(\vec{x},-\Omega)\alpha_{out}(-\vec{x})|^2\;,
                                      \label{ispectrum2}
\end{eqnarray}
where
\begin{eqnarray}\label{SN-}
SN_- = 
\eta \int_{R_1+R_2}\!\!\!\! d\vec{x}~
   |\alpha_{out}(\vec{x})|^2 
\end{eqnarray}
and we used Eq.(\ref{aout}). By comparing 
with Eqs.(\ref{vpm}) and (\ref{SNLO}), we see 
that this expression coincides with $V^{(-)}_{\phi_L}(\Omega)$ 
if we take:
\begin{eqnarray}
\alpha_L(\vec{x})=\alpha_{out}(\vec{x})\;.
\end{eqnarray}
This is expected because a LO with the 
configuration of the output field just picks up the 
amplitude fluctuations. The relation becomes 
even more precise in the phase-sensitive 
case $\alpha_{in}(\vec{x})=\alpha_{in}(-\vec{x})$. 
In this case, assuming $\eta=1$ and that the 
pixel area is small with respect to $x_0^2$ and 
to the square of the scale of variation of 
$\alpha_{out}(\vec{x})$, one has:  
\begin{eqnarray}
{V_{i_-}(\Omega) \over SN_-}=
\tilde F (\vec{x},\Omega)
=   |\overline{u}(\vec{x},\Omega)e^{-i\phi_{out}(\vec{x})}
    -\overline{v}^*(\vec{x},-\Omega)e^{i\phi_{out}(\vec{x})}|^2\;,
                                    \label{ispectrum3}
\end{eqnarray}
where we set $\alpha_{out}(\vec{x})=
\rho_{out}(\vec{x}) \exp(i\phi_{out}(\vec{x}))$, and 
$\vec{x}$ is the central point of any of the two 
symmetrical pixels. The result coincides with that 
of Eq.(\ref{vpm2}) where $\phi_L$ is replaced by 
$\phi_{out}$.
The link between intensity fluctuations and 
quadrature fluctuations allows now to 
analyse immediately the case of phase 
fluctuations, which coincide with the quadrature
fluctuations obtained by using a LO which 
displays a phase shift of $\pi / 2$ with 
respect to LO which provides the amplitude 
fluctuations. 
Therefore, in terms of pixels, we are lead 
consider the observables $Z^{(j)}_{\phi_L+ {\pi / 2}}(t)$
(see Eq.(\ref{Zx})) with 
$\phi_L=\phi_{out}$. 
This naturally induces us to focus on the 
observable $
Z^{(+)}(t)=Z^{(1)}_{\phi_{out}+{\pi / 2}}(t)+
Z^{(2)}_{\phi_{out}+{\pi / 2}}(t)$,
see Eq.(\ref{Zxpm}),
which measures the degree of anticorrelation 
between the phase fluctuations in the two 
symmetrical pixels 1 and 2. The spectrum 
$V^{(+)}_{\phi_L+\pi/2}(\Omega)$ coincides with 
$V^{(-)}_{\phi_L}(\Omega)$, which as we have 
seen is identical to $V_{i-}(\Omega)$ given by 
Eq.(\ref{ispectrum3}). Therefore for large 
amplification the fluctuations of 
$Z^{(+)}(t)$ are well below the shot noise 
level, which implies that the phase fluctuations 
in the two symmetrical pixels are strongly 
anticorrelated, exactly as the amplitude 
fluctuations are strongly correlated.

\section{Conclusion}

In this article we analyzed extensively a system 
formed by an optical parametric amplifier with 
some imaging lenses. Amplification of optical 
images by OPA has been already studied 
in the literature \cite{Devand}, but only 
from a classical viewpoint. 

Our results hold both for a phase-sensitive 
configuration (symmetrical input image) and 
for a phase insensitive one (asymmetrical injection).

We demonstrated that the two output twin images 
exhibit a complete spatial EPR entanglement. 
This was shown, first of all, by considering 
a pair of orthogonal quadrature components 
of the output field. In the case of 
local oscillator symmetrical with respect to 
the system axis, we found a precise 
prescription for the phase $\phi_L$ of the 
local oscillator (see introduction) in order 
to observe maximal correlation between 
symmetrical pixels of the two output images. 
The optimal value for the phase is that which 
corresponds to the amplitude fluctuations of 
the output images in the phase sensitive 
configuration, when the phase of the symmetrical 
input images is selected to obtain maximal 
amplification. 

We have shown also that a performance very close 
to that of the ideal case of the optimal LO phase 
can be obtained in practice by using a LO with 
a quadrature wavefront (as one has in gaussian beams) 
with the curvature used as optimisation parameter. 

The connection between quadrature fluctuations and 
amplitude/phase fluctuations has in turn allowed us 
to conclude also that, while intensity 
fluctuations are strongly correlated in the twin 
images, phase fluctuations are strongly 
anticorrelated in the same amount. An amusing 
analogy with amplitude and phase fluctuations 
in entangled twin images is provided by a 
fossile broken in two pieces (see Fig.3). 
We see that the structures in the two pieces 
have the same ``amplitude'' = thickness, but 
opposite ``phase'' (one is concave and the other 
convex, one is righthanded and the other 
lefthanded). 

It is important to underline that, while the 
results for intensity and phase fluctuations 
hold only in presence of an input image, the 
result on EPR entanglement of quadrature 
components hold also in absence of any input 
image, i.e. in the case of the pure parametric 
down-conversion as in \cite{fluor}. This is 
important for the applications to 
quantum teleportation of optical images 
\cite{Sokolov}, as a generalisation of the 
Braunstein-Kimble \cite{braunstein,Vaidman} scheme 
for a single mode field, or to quantum 
cryptography with images. 

We observe finally that our results hold 
also if the OPA is replaced by an optical 
parametric oscillator below threshold with 
plane mirrors (see \cite{Images} in this 
connection). As a matter of fact, also in this 
case one has an input-output relation of the 
form (\ref{inputoutput2}), and the results are based only on this 
relation and on the general properties 
of the functions $\overline{u}$ and $\overline{v}$. 

ACKNOWLEDGEMENTS:

The authors warmly thank
M.I. Kolobov for usefull remarks and criticisms.
This work was supported by the network QSTRUCT of the 
TMR programme of the EU.

\section{Figure captions}
% FIGURE 1 %
\begin{figure}[p]
\label{fig1}
\caption[]{Schematic diagram of the parametric image amplifier. The two-lens
telescopic system allows to obtain two amplified copies of the input image
that are strongly quantum correlated to each other,
thereby the name twin images. The device is phase-sensitive
when the input image is symmetrical, phase-insensitive when it
is confined in the upper half of plane $P_1$. 
$f$ is the focal distance of the lenses.}
\end{figure}

\begin{figure}[p]
\label{nred}
\caption[]{Plot of the noise reduction factor $F(\vec{x},0)$.
Subscripts (a) refers to the optimal phase of the LO
while (b) and (c) refer respectively to 
a constant phase and to a phase with a quadratic dependence 
on the distance from the optical axis.
The dashed line is the phase-sensitive gain of the OPA 
(see Eq.(\ref{Gain})
divided by a factor 10). $\Delta_0 l_c=0.5$ and $|\sigma|l_c=1.5$.}
\end{figure}

\begin{figure}[p]
\label{fossile}
\caption[]{
Analogy between a broken fossile and 
quantum entangled images (see text). 
}
\end{figure}

\begin{thebibliography}{99}
\bibitem{Jost} R. Jost, Optical Express (1998) 
(same special issue as ours); new article of Salch et al., 
Phys. Rev. A {\bf 2}, 196 (Spetember 2000).
\bibitem{Images} A. Gatti, E. Brambilla, L.A. Lugiato \& 
		M.I. Kolobov, Phys. Rev. Lett. {\bf 83}, 1763 (1999).
\bibitem{Fabre}
C. Fabre et al., \dots
\bibitem{Images2} A. Gatti, E. Brambilla, L.A. Lugiato \&  
	M.I. Kolobov, J.~Opt. B: Quantum Semiclass.~ Opt., in press (2000).
\bibitem{epr35}A.~Einstein, B.~Podolsky, and N.~Rosen, Phys.~Rev.~{\bf 47},
               777 (1935).
\bibitem{epr92}M.~D.~Reid, Phys.~Rev. A {\bf 40}, 913
                (1989); Z.~Y.~Ou, S.~F.~Pereira and H.~J.~Kimble, Appl.~Phys.~B,                  {\bf 55}, 265 (1992).
\bibitem{epr}A. Gatti, L.A. Lugiato, K.I. Petsas \& I. Marzoli,
	Europhys. Lett. {\bf 46}, 461 (1991). Anche Opt. Com. in onore 
di Scully
\bibitem{kolobov99} I.~V.~Sokolov, M.~I.~Kolobov and L.~A.~Lugiato,
                   Phys.~Rev.~A {\bf 60},~2420 (1999).
\bibitem{kolobov89} M.~I.~Kolobov and I.~V.~Sokolov, Sov.~Phys.~JETP {\bf 69},~1097 (1989);
	Phys.~Lett.~A {\bf 140},~101 (1989).
\bibitem{kolobov95}  M.~I.~Kolobov and L.~A.~Lugiato, Phys.~Rev A {\bf 52}, 5930 
                (1995).
\bibitem{Levenson} J.A. Levenson, I. Abram, T. Rivera, P. Fayolle,
J.C. Garreau and Ph. Grangier, Phys. Rev. Lett. {\bf 70}, 270 (1998).
\bibitem{Kumar} P. Kumar
\bibitem{caves} C.M. Caves, Phys. Rev. D {\bf 26}, 1817 (1982)
\bibitem{fluor} E. Brambilla, A.Gatti and L.A. Lugiato, 
to be submitted for publication.
\bibitem{Devand} See e.g. F. Devaud and F. Lantz, J. Opt. Soc. 
Am. B {\bf 12}, 2245 (1995).
\bibitem{Sokolov}
I.V. Sokolov, M.I. Kolobov, A. Gatti and L.A. Lugiato, 
e-print quant-ph/0007026.
\bibitem{braunstein}
S.~L.~Braunstein and H.~J.~Kimble, Phys.~Rev.~Lett. {\bf 80}, 869 (1998);
A. Furusawa, J. L. Sorensen, S. L. Braunstein, C. A. Fuchs, H. J. Kimble, and E. S. Polzik
       Science {\bf 282}, 706-709 (1998).
\bibitem{Vaidman} Vaidman, ...
\end{thebibliography}
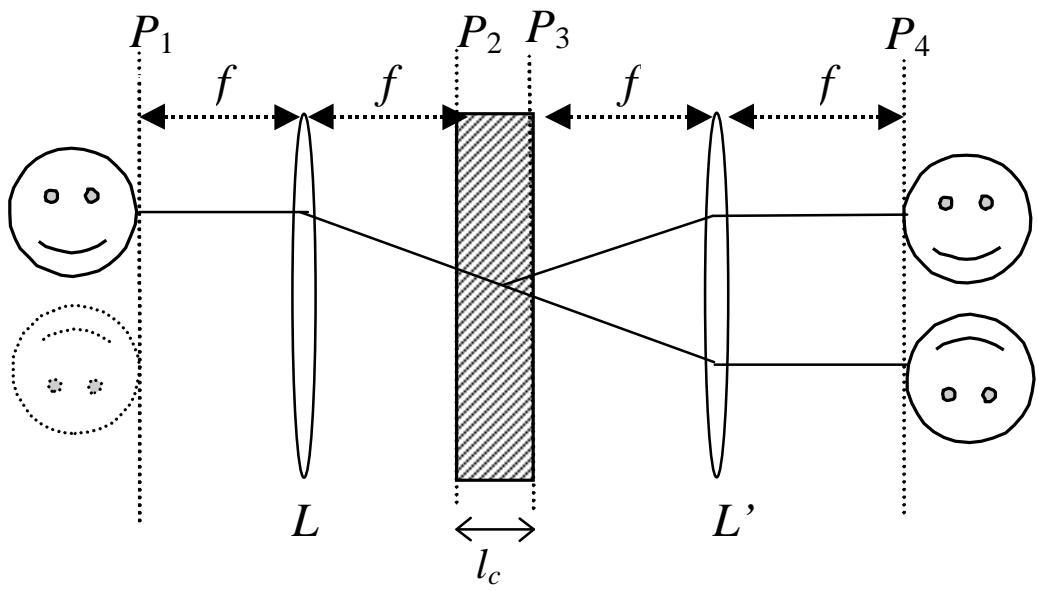
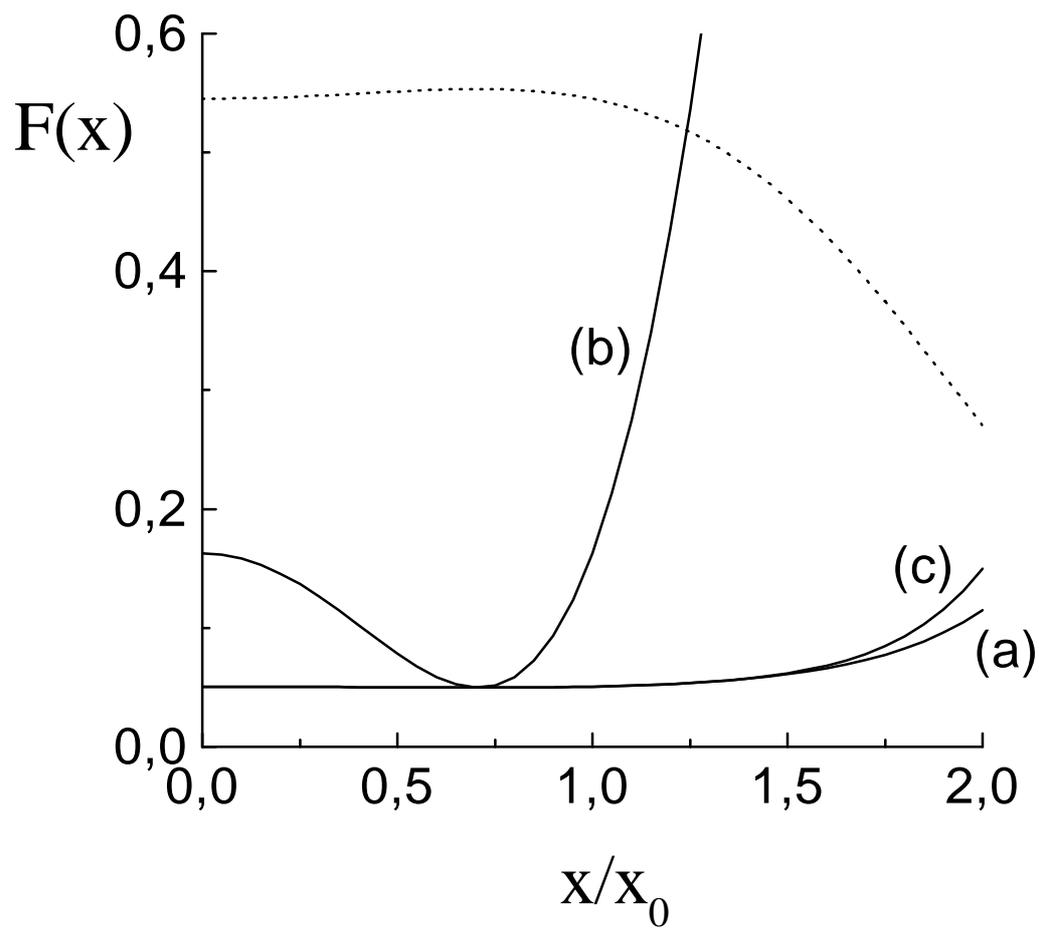
\end{document}